\documentclass[pra,twocolumn,letterpaper]{revtex4}

\usepackage{amssymb,amsmath,amsthm,amsfonts}
\usepackage{graphicx}
\usepackage{epsfig}
\usepackage{subfigure}

\newcommand{\field}[1]{\mathbb{#1}}

\newcommand{\ket}[1]{\ensuremath{| #1 \rangle}}
\newcommand{\bra}[1]{\ensuremath{\langle #1 |}}
\newcommand{\ketbra}[2]{\ensuremath{| #1 \rangle \langle #2 |}}
\newcommand{\braket}[2]{\ensuremath{\langle #1 | #2 \rangle}}
\newcommand{\set}[1]{\ensuremath{\left \{ #1 \right \}}}
\newcommand{\coltwo}[2]{\ensuremath{\left[ \begin{array}{c} #1 \\ #2 \end{array} \right]}}

\newcommand{\myset}[2]{\ensuremath{\left \{ \left. ~ #1 ~ \right | ~ #2 ~ \right \}}}
\newcommand{\mtxtwotwo}[4]{\ensuremath{\left[ \begin{array}{cc} #1 & #2 \\ #3 & #4 \end{array} \right]}}

\newcommand{\brackets}[1]{\ensuremath{\left( #1 \right)}}
\newcommand{\pauli}[1]{\ensuremath{\mathcal{G}_{#1}}}
\newcommand{\twiddle}[1]{\ensuremath{\tilde{#1}}}

\newcommand{\bs}[1]{\ensuremath{\set{0,1}^#1}}
\newcommand{\mbs}[1]{\ensuremath{\mathbf{#1}}}

\newtheorem{thm}{Theorem}[section]
\newtheorem{cor}[thm]{Corollary}
\newtheorem{lem}[thm]{Lemma}

\theoremstyle{remark}
\newtheorem{rem}[thm]{Remark}

\theoremstyle{definition}
\newtheorem{exam}[thm]{Example}

\begin{document}

\markboth{Adam G. D'Souza, Jop Bri\"et and David L. Feder}
{Testing equivalence of pure quantum states and graph states under SLOCC}

\title{Testing equivalence of pure quantum states and graph states under SLOCC}

\author{Adam~G.~D'Souza$^1$, Jop~Bri\" et$^2$, and David~L.~Feder$^{1}$} 
\affiliation{$^1$Department of Physics and Astronomy and Institute for Quantum
Information Science, University of Calgary, Calgary, Alberta, Canada, T2N 1N4\\
$^2$Quantum Computing and Advanced Systems Research,
Centrum Wiskunde \& Informatica, P.O. Box 94079
NL-1090 GB, Amsterdam, The Netherlands}

\date{\today}

\begin{abstract}
A set of necessary and sufficient conditions are derived for the equivalence 
of an arbitrary pure state and a graph state on $n$ qubits under stochastic 
local operations and classical communication (SLOCC), using the stabilizer 
formalism. Because all stabilizer states are equivalent to a graph state by local
unitary transformations, these conditions constitute a classical algorithm
for the determination of SLOCC-equivalence of pure states and stabilizer states.
This algorithm provides a distinct advantage over the direct solution of the
SLOCC-equivalence condition $\ket{\psi} = S \ket{g}$ for an unknown invertible local
operator $S$, as it usually allows for easy§ detection of states that are not 
SLOCC-equivalent to graph states.
\end{abstract}


\maketitle

%%%%%%%%%%%%%%%%%%%%%%%%%%%%%%%%%%%%%%%%%%%%%%%

\section{Introduction}
\label{sec:introduction}

\subsection{Background and Motivations}

Graph states are the quantum analogs of classical graphs, in which qubits
correspond to vertices and maximal two-qubit entanglement is represented by an 
edge connecting the vertices. These highly entangled states of many qubits have 
been the subject of much theoretical study over the past several 
years~\cite{Hein2006,Briegel2009}, and have been physically implemented by 
several experimental groups using both nuclear magnetic resonance~\cite{Ju2008}
and photons~\cite{Kiesel2005,Vallone2007,Lu2007,Lu2008,Tokunaga2008}.
The interest is due in large part to the fact that with successive 
measurements of the constituent qubits of certain graphs, together with 
classical processing, it is possible to perform arbitrary quantum algorithms. 
In this sense, graph states serve as `universal resources' for 
measurement-based quantum computation 
(MBQC)~\cite{Gross2007a,Gross2007,VandenNest2006,VandenNest2007}.

Graph states have several additional applications for quantum information
processing. They are central to the theory of quantum error 
correction~\cite{Gottesman1997}, because all stabilizer code states are 
equivalent to graph states under local unitary transformations, in fact under 
a very restricted set of operators belonging to the local Clifford 
group~\cite{Schlingemann2002}. In recent years, it has been shown that graph 
states can be used for multiparty quantum secret sharing~\cite{Markham2008},
are closely related to classical spin models~\cite{Hubener2008,Cuevas2008},
and are associated with undecidable logic theories~\cite{VandenNest2008}.
When the graph states are suitably 
weighted~\cite{Dur2005,Calsamiglia2005,Anders2006,Anders2007,Gross2007}, they 
can efficiently approximate the ground states of strongly interacting spin
systems~\cite{Anders2006,Anders2007}, and can be used in order to implement 
random circuits using only measurements~\cite{Plato2008}. 

While graph states have wide uses in quantum information theory, it is not 
clear what other kinds of states can be used to accomplish the same kinds of 
tasks, nor even what are the essential properties of graph states that make
them so useful. One might na\" \i vely expect that any multipartite state with
a large amount of entanglement (carefully defined in this multi-party context)
would be equally useful. Very recently, however, it was shown independently
by two groups that the number of states that can serve as universal resources 
for MBQC decreases exponentially with the total number of 
qubits~\cite{Gross2008,Bremner2008}. Given a quantum pure state with $n$ qubits, 
it would be useful to have a set of criteria such that one would be able to determine if it were in fact a 
universal resource for MBQC, and by extension a useful resource for more
general quantum information tasks.

The goal of the present work is to partially address this issue by considering
$n$-qubit pure states $|\psi\rangle$ that are equivalent to 
$n$-qubit graph states $|g\rangle$ by stochastic local operations and 
classical communication (SLOCC)~\cite{Bennett1996,Vidal1999,Duer2000}.
Mathematically, the two states are connected by a tensor product of $n$ local operations, 
$|\psi\rangle=S|g\rangle$, where $S=S_1\otimes S_2\otimes\cdots\otimes S_n$
and $S_i\in\text{GL}(2,\field{C})$. In principle, any state that is 
SLOCC-equivalent to a graph state can accomplish the same tasks as the 
original graph state with finite probability: one first performs the inverse
operations $S^{-1}|\psi\rangle$ and then works with the resulting graph state. 
Implementing the generally non-unitary operators $S_i^{-1}$ is difficult in 
practice, however. Rather, one would perform the conversion with non-zero 
probability by (deterministic) LOCC using an appropriate positive operator 
valued measure (POVM) on each qubit~\cite{Lo2001}; Neumark's theorem describes 
how to translate a POVM into a projective measurement~\cite{Peres1990}. 
Alternatively, the desired quantum information task would be accomplished 
directly using some set of POVMs.

It is worth noting that the probability of success of an information processing task on a SLOCC-transformed resource state is often 
much lower than that corresponding to the same task on the original resource. For instance, the probability of
successfully performing MBQC on a SLOCC-transformed cluster state, a well-known resource for MBQC, will be exponentially suppressed in the
number of gates appearing in the computation, since each gate will be probabilistic in general. This is in contrast to the case of MBQC on a perfect
cluster state, which is deterministic. In order to determine the true utility of a SLOCC-transformed resource state for MBQC, one could adopt a strategy, such as a
percolation strategy~\cite{Kieling2007,Browne2008}, for distilling a perfect resource from the SLOCC-transformed one, on which MBQC would be deterministic. 
If the reduction in size of the SLOCC-transformed resource scales polynomially in the original size, then we could reasonably say
that the SLOCC-transformed resource state is a useful resource for universal MBQC itself.

The classification of quantum pure states under SLOCC transformations has been
the subject of much study in recent years. On two qubits, it is known that all entangled
pure states are asymptotically SLOCC-equivalent~\cite{Bennett1996}. On three 
qubits there are precisely two inequivalent classes~\cite{Duer2000}, 
represented by the GHZ state~\cite{Greenberger2007} and the W 
state~\cite{Coffman2000}. The determination of the number of 
SLOCC-inequivalent classes on four qubits remains controversial. Early work 
indicated that there are at least nine classes~\cite{Verstraete2002}; more 
recently, it was shown that there are at least 28 truly inequivalent 
classes~\cite{Li2007,Li2007a}, of which all but eight are 
non-degenerate~\cite{Lamata2007}. Even the determination of bipartite 
separability of quantum pure states, let alone the classification of quantum 
pure states into multipartite SLOCC classes, is an NP-hard 
problem~\cite{Gurvits2003} .

Given the challenges of classifying quantum pure states under SLOCC even for
small numbers of qubits, one might assume that determining SLOCC-equivalence
between an arbitrary $n$-qubit state $|\psi\rangle$ and a graph state would be 
difficult. While this seems to be true, there does exist a 
classical test of the SLOCC-equivalence of these two states that is generally efficient at detecting SLOCC-inequivalence. The result hinges on the 
stabilizer formalism for graph states.
By implication, any state that is SLOCC-equivalent to a graph state must also have a separable
stabilizer, though in general consisting of separable non-unitary and 
non-Hermitian operators. Such generalized stabilizers were recently considered
in a different context~\cite{Plenio2007}. A related result of the present
work is an algorithm for obtaining a separable stabilizer for an arbitrary
pure quantum state, if one exists. The manuscript is organized as follows. Section~\ref{sec:notation} gives some important definitions and notation. Section~\ref{sec:mainthm} states the main result of this work, Theorem~\ref{mainthm}, together with some examples of how to apply it. Section~\ref{sec:derivation} gives its full derivation. Section~\ref{sec:test} explains how one would go about using the theorem to actually test for SLOCC-equivalence between a graph state and a pure state. In Section~\ref{sec:constructing}, a related result describing how one can build the separable stabilizer (in the generalized sense) of a SLOCC-transformed state is provided. Finally, Section~\ref{sec:conclusion} summarizes the conclusions of this work.    

\section{Definitions and Notation}
\label{sec:notation}

In this paper, we use the following notational conventions.

\begin{itemize}
\item \textbf{Pauli operators.} For integer $k \geq 1$, we define the set of $k$-qubit operators $\pauli{k} := \alpha \set{I,X,Y,Z}^{\otimes k}$ for
$\alpha \in \set{\pm 1,\pm i}$ (i.e., the set of $n$-qubit Pauli operators modulo factors of fourth roots of unity).
\item \textbf{Set notation for targets of separable quantum operations.} The set of qubits on which a separable multi-qubit operator acts is written as a subscript on the operator. For example, the symbol $O_\mathcal{J}$ denotes an operator $O$ acting on a set of qubits $\mathcal{J} \subseteq \mathcal{V}$, where $\mathcal{V}$ is the set of all qubits. Boldface letters such as \mbs{j} denote bit strings, and \mbs{j(0)} and \mbs{j(1)} respectively denote the set of entries of \mbs{j} that are equal to 0 and 1. Thus, $\mathcal{O}_{\mbs{j(0)}}$ is an operator acting on all qubits whose indices correspond to the positions of the bits of \mbs{j} that are equal to $0$, and similarly for $\mathcal{O}_{\mbs{j(1)}}$.
\item \textbf{Graph state.} The $n$-qubit graph state $\ket{g}$ corresponding to an $n$-vertex graph $G=\brackets{\mathcal{V},\mathcal{E}}$, where $\mathcal{V}$ and $\mathcal{E}$ respectively denote the sets of vertices and edges of $G$, is the stabilizer state that is the unique simultaneous eigenstate of the operators $\sigma_i = X_iZ_{\mathcal{N}\brackets{i}}$, where $\mathcal{N}\brackets{i}$ is the neighbourhood of the vertex $i$. Note that all stabilizer states are equivalent to a graph state under local Clifford operations~\cite{Schlingemann2002}. Thus, the results presented in this paper in the context of graph states are generalizable to arbitrary stabilizer states.
\item \textbf{SLOCC operators and SLOCC-equivalence.} The set of all operators corresponding to a branch of a SLOCC protocol on an $n$-qubit state is given by
\begin{equation}
\nonumber \rm{SLOCC} \brackets{2^{\it{n}}} := \myset{\it{S} = \displaystyle \bigotimes_{i=\rm{0}}^{n-\rm{1}} S_i}{\it{S_i} \in \rm{GL}\brackets{2, \mathbb{C}}} \text{.}
\end{equation}
Two $n$-qubit quantum pure states $\ket{\psi}$ and $\ket{g}$ are said to be SLOCC-equivalent if and only if there exists an operator $S \in \rm{SLOCC} \brackets{2^{\it{n}}}$ such that
\begin{equation}
\nonumber \ket{\psi} = S \ket{g} \text{.}
\end{equation}
\end{itemize}

\section{Main Theorem and its Intuitive Justification}
\label{sec:mainthm}
The main theorem of the paper is the following:

\begin{thm}
Let \ket{g} be an $n$-qubit graph state with underlying graph $G=\brackets{\mathcal{V},\mathcal{E}}$, stabilized by $\Sigma(G)=\myset{\sigma_i}{\sigma_i \in \pauli{n}, \sigma_i \ket{g} = \ket{g}}$. Let $S \in \rm{SLOCC} \brackets{2^{\it{n}}}$. Then, any $n$-qubit pure state \ket{\psi} satisfies the conditions
\begin{equation}
\bra{\psi^*}Y_\mathcal{V} S Y_\mathcal{V} \sigma_i Z_\mbs{j(1)} S^{-1} \ket{\psi}=\det\left(S\right) \delta_{j,0},
\label{cond1}
\end{equation}
for all $i \in \mathcal{V}$ and $\mbs{j} \in \bs{n}$ if and only if $\ket{\psi} = S \ket{g}$.
\label{mainthm}
\end{thm}

In this Theorem, \bra{\psi^*} refers to the complex conjugate of \bra{\psi} in the computational basis, or equivalently the transpose of \ket{\psi} in the computational basis. The operator $Z_{\mbs{j(1)}}$ describes Pauli $Z$ operators acting on all qubits in positions where the bit string $\mbs{j}$ has entries equal to 1, and identity operators on all other qubits. Note that \ket{\psi} need not be normalized; however, it can be assumed to be so without loss of generality, as an unnormalized state \ket{\psi} is SLOCC-equivalent to \ket{g} if and only if its normalized counterpart $c \ket{\psi}$ (where $c \in \mathbb{C}$) is as well. While the full details of the proof of this theorem are deferred until Section~\ref{sec:derivation}, the theorem can be justified intuitively as follows. The graph state $\ket{g}$
has a separable Pauli stabilizer $\sigma_i\ket{g}=\ket{g}$ with elements 
$\sigma_i=\bigotimes_{j=0}^{n-1}\sigma_{ij}$ with the $\sigma_{ij}\in\pauli{1}$
consisting of Pauli matrices $X$, $Y$, or $Z$. Suppose that $\ket{\psi}$ is in 
fact SLOCC-equivalent to the graph state $\ket{g}$, i.e.\ that 
$\ket{\psi}=S\ket{g}$ with $S=\bigotimes_{j=0}^{n-1}S_j$ and 
$S_j\in GL(2,\field{C})$. Because the $S_j$ transform each of the Pauli
operators in the graph-state stabilizer, the state $\ket{\psi}$ must also have 
a separable {\it generalized} stabilizer, but now consisting in general of
non-Hermitian operators $\tilde{\sigma}_{ij}=S_j\sigma_{ij}S_j^{-1}$ (note
that in general $S_j^{-1}\neq S_j^{\dag}$). Just as single-qubit Pauli errors
project stabilizer states out of their code space, likewise the application of
any number of transformed Pauli operators on the state $\ket{\psi}$ yield 
states that are orthogonal. Thus the generalized orthogonality 
conditions~(\ref{cond1}).

\subsection{Additional Considerations on the Main Theorem}
\label{sec:considerations}
While in principle all of the $4^n$ conditions obtained from the $2^n$ choices of stabilizer element and the $2^n$ choices of \mbs{j} in Eq.~(\ref{cond1}) are necessary for SLOCC-equivalence of \ket{\psi} to \ket{g}, in practice some of them do not constrain the form of the supposed SLOCC operator $S$ connecting \ket{\psi} and \ket{g}, and are thus not useful for SLOCC-equivalence testing. We can completely partition the conditions from Eq.~(\ref{cond1}) into three disjoint categories:
\begin{enumerate}
\item \textbf{Category I}. These are the conditions of the form of Eq.~(\ref{cond1}) where $(-1)^n \neq (-1)^{W}$, where $W$ is the weight of the unknown operator $S Y_\mathcal{V} \sigma_i Z_\mbs{j(1)} S^{-1}$, i.e. the number of sites on which the local operator acting is not proportional to the identity.
\item \textbf{Category II}. These are those sets of conditions not falling into Category I that have the form
\begin{equation}
\label{formofconstraints} \bra{\psi^*} Y_{\mathcal{V}} \mathcal{O}_{\mathcal{J}}^{(i)} \ket{\psi} = 0 \\
\end{equation}
where $\set{\mathcal{O}_{\mathcal{J}}^{(i)}}$ is the set of operations involved in the SLOCC-equivalence conditions~(\ref{cond1}) that have target $\mathcal{J}$.
\item \textbf{Category III.} These are conditions of the form 
\begin{equation}
\label{formofconstraints} \bra{\psi^*} Y_{\mathcal{V}} \mathcal{O}_{\mathcal{J}}^{(i)} \ket{\psi} = A_i \\
\end{equation}
where $A_i \neq 0$ for at least one $i \in \set{0,1,\dots,|\mathcal{J}|-1}$.
\end{enumerate}
Category I conditions are automatically satisfied for all pure states \ket{\psi}, not just states that are SLOCC-equivalent to a graph state. Similarly, Category II conditions are automatically satisfied by any pure state \ket{\psi} that is SLOCC-equivalent to the desired graph state and can also be omitted when performing the inequivalence test. Thus, the only conditions that need to be tested for a complete SLOCC-equivalence test are Category III conditions. These statements are formally made as follows:
\begin{lem}
\label{oddevenlem}
(Category I conditions are automatically satisfied.) Let \ket{\psi} be an $n$-qubit pure state defined on the set of qubits $\mathcal{V}$. Let $J \subseteq \mathcal{V}$. Let $\twiddle{Z}_J = SZ_JS^{-1}$ for some invertible $S$. Then, $\bra{\psi^*} Y_\mathcal{V} \twiddle{Z}_J \ket{\psi} = 0$ if $|\mathcal{V} \setminus J|$ is odd.
\end{lem}
\begin{lem}
\label{allzeroslemma}
(Category II conditions are automatically satisfied.) Let \ket{\psi} be an $n$-qubit pure state that is SLOCC-equivalent to an $n$-qubit graph state. Then the conditions of the form
\begin{equation}
\label{allzeroseq} \bra{\psi^*} \brackets{Y_{\mathcal{V}}} S \mathcal{O}^{(i)}_{\mathcal{J}} S^{-1} \ket{\psi} = 0
\end{equation}
where $\set{O^{(i)}_{\mathcal{J}}} \subseteq \pauli{\left| \mathcal{J} \right|}$ is the set of all $n$-qubit Pauli operators with target $\mathcal{J}$, are satisfied independent of the choice of $S$.
\end{lem}
The proofs of Lemma~\ref{oddevenlem} and Lemma~\ref{allzeroslemma} are given in Section~\ref{sec:test}. 

\subsection{Examples}
Theorem \ref{mainthm} gives a set of necessary conditions for SLOCC-equivalence of an $n$-qubit pure state \ket{\psi} to an $n$-qubit graph state \ket{g}. Each of these conditions contains between $0$ and $n$ unknown transformed Pauli operators, which will be denoted $\twiddle{X}$, $\twiddle{Y}$ or $\twiddle{Z}$, where $\twiddle{Z} = SZS^{-1}$ for some invertible $S$, and similarly for \twiddle{X} and \twiddle{Y}. For each choice of stabilizer element $\sigma_i$, the number of unknown local operators appearing is bounded below by $|\sigma_i(I)| + |\sigma_i(Z)|$, the number of sites on which $\sigma_i$ acts locally as either $I$ or $Z$. The condition for which no unknown local operators appear is obtained from the unique choice for $\sigma$ where $|\sigma(I)| + |\sigma(Z)| = 0$, or equivalently where $\sigma$ acts locally on all sites as either $X$ or $Y$. This stabilizer element results from multiplying all $n$ generators together, because for each qubit the $X$ multiplies either an even or odd number of $Z$ operators. Therefore, the local operators composing $Y_{\mathcal{V}} \sigma \in \pauli{n}$ are all proportional either to $I$ or to $Z$. Moreover, the number of unknown local operators appearing in Eq. (\ref{cond1}) can take any value from $0$ to $n$. Similarly, consider the case in which $\sigma$ is the product of all but one of the generators. The number of unknown local operators appearing in conditions resulting from this choice can be anywhere from 1 to $n$. In general, if the chosen stabilizer element is a product of all but $k$ of the generators, then at least $k$ unknown local operators will appear in the associated conditions.
\begin{exam}
\label{threequbitsexample}
Suppose we would like to test if an arbitrary pure quantum state \ket{\psi} is
SLOCC equivalent to the three-qubit linear cluster state \ket{g_3}, which is
stabilized by the generators $\{\sigma_1,\sigma_2,\sigma_3\}
=\{X\otimes Z\otimes I,Z\otimes X\otimes Z,
I\otimes Z\otimes X\}$. Let's rewrite these as 
$\{\sigma_1,\sigma_2,\sigma_3\}=\{XZI,ZXZ,IZX\}$ for brevity, with the position 
of the Pauli or identity operator representing the qubit on which it acts.
The remaining non-trivial stabilizers are generated by multiples of these: 
$\{\sigma_4,\sigma_5,\sigma_6,\sigma_7\}=\{YYZ,XIX,ZYY,-YXY\}$. Eq.~(\ref{cond1}) with the choices $\sigma = \sigma_7$ (the unique stabilizer element for which $|\sigma(I)| + |\sigma(Z)| = 0$) and $\mbs{j} = \mbs{010}$ gives
\begin{equation}
(\bra{\psi^*} Y_{\mathcal{V}}) \ket{\psi}=0, \label{q0cond} \\
\end{equation}
in which no unknown local operators appear. Eq.~(\ref{cond1}), together with the stabilizers 
$\sigma_4$ through $\sigma_7$ (the elements for which $|\sigma(I)| + |\sigma(Z)| \leq 1$) and the appropriate choices of $\mbs{j}$, yields the conditions
\begin{eqnarray}
(\bra{\psi^*} Y_{\mathcal{V}}) II\tilde{X}\ket{\psi}&=&-i\;\det(S); \label{q1cond1} \\
(\bra{\psi^*} Y_{\mathcal{V}}) II\tilde{Y}\ket{\psi}&=&0; \\
(\bra{\psi^*} Y_{\mathcal{V}}) II\tilde{Z}\ket{\psi}&=&0; \label{q1cond3} \\
(\bra{\psi^*} Y_{\mathcal{V}}) I\tilde{X}I\ket{\psi}&=&0; \label{q2cond1} \\
(\bra{\psi^*} Y_{\mathcal{V}}) I\tilde{Y}I\ket{\psi}&=&0; \\
(\bra{\psi^*} Y_{\mathcal{V}}) I\tilde{Z}I\ket{\psi}&=&-i\,\det(S); \label{q2cond3} \\
(\bra{\psi^*} Y_{\mathcal{V}}) \tilde{X}II\ket{\psi}&=&-i\,\det(S); \label{q3cond1} \\
(\bra{\psi^*} Y_{\mathcal{V}}) \tilde{Y}II\ket{\psi}&=&0; \\
(\bra{\psi^*} Y_{\mathcal{V}}) \tilde{Z}II\ket{\psi}&=&0.\label{q3cond3}
\end{eqnarray}
Note that this particular graph state is symmetric under reflection about 
qubit 2, so that conditions~(\ref{q3cond1}-\ref{q3cond3}) are the symmetry 
counterparts of (\ref{q1cond1}-\ref{q1cond3}). Eq.~(\ref{q0cond}) is a Category I condition. The groups of conditions 
(\ref{q1cond1}-\ref{q1cond3}), (\ref{q2cond1}-\ref{q2cond3}), and 
(\ref{q3cond1}-\ref{q3cond3}) each correspond to transformations on a single 
qubit and are all Category III conditions. Similarly, conditions corresponding to transformations on two qubits can be obtained from all choices of $\sigma$ except for $\sigma_0 = III$, and those corresponding to transformations on three qubits can be obtained from any choice of $\sigma$.
\end{exam}

\begin{exam}
Consider now the five-qubit cluster state, stabilized by the generators $\set{\sigma_1,\sigma_2,\sigma_3,\sigma_4,\sigma_5 } = \set{XZIII, ZXZII, IZXZI, IIZXZ, IIIZX}$. Choosing the stabilizer element $\sigma_6 = \sigma_1 \sigma_2 \sigma_3 \sigma_4 = YXXYZ$ and $\mbs{j} = \mbs{01100}$ and $\mbs{j}=\mbs{01101}$ respectively yields
\begin{eqnarray}
\label{example2X} \bra{\psi^*}Y_{\mathcal{V}}IIII\twiddle{X} \ket{\psi} & = & 0; \\
\label{example2Y} \bra{\psi^*}Y_{\mathcal{V}}IIII\twiddle{Y} \ket{\psi} & = & 0.
\end{eqnarray}
Furthermore, choosing the stabilizer element $\sigma_7 = \sigma_1 \sigma_2 \sigma_3 \sigma_4 \sigma_5 = -YXXXY$ and $\mbs{j} = \mbs{01111}$ gives
\begin{equation}
\label{example2Z} \bra{\psi^*}Y_{\mathcal{V}}IIII\twiddle{Z} \ket{\psi} = 0.
\end{equation} 

Eqs.~(\ref{example2X}-\ref{example2Z}) are Category II conditions, and are thus satisfied by any $n$-qubit pure state \ket{\psi} that is SLOCC-equivalent to the five-qubit cluster; these conditions do not constrain the unknown elements of \twiddle{X}, \twiddle{Y} or \twiddle{Z}. On the other hand, consider the stabilizer element $\sigma_8 = \sigma_1 \sigma_2 \sigma_4 \sigma_5 = YYIYY$. The choices $\mbs{j} = \mbs{00000}$ and $\mbs{j} = \mbs{00100}$ respectively yield
\begin{eqnarray}
\label{example2Y1} \bra{\psi^*}Y_{\mathcal{V}}II\twiddle{Y}II \ket{\psi} & = & \det \brackets{S}; \\
\label{example2X1} \bra{\psi^*}Y_{\mathcal{V}}II\twiddle{X}II \ket{\psi} & = & 0.
\end{eqnarray}
\end{exam}
Next, choose the stabilizer element $\sigma_7$ and $\mbs{j} = \mbs{01010}$ to obtain
\begin{equation}
\label{example2Z1} \bra{\psi^*}Y_{\mathcal{V}}II\twiddle{Z}II \ket{\psi} = 0.
\end{equation} 
Eqs.~(\ref{example2Y1}-\ref{example2Z1}) are Category III conditions, and do impose constraints on the unknowns appearing in the elements of \twiddle{X}, \twiddle{Y} and \twiddle{Z}.

\section{Derivation of Necessary and Sufficient Conditions}
\label{sec:derivation}
By definition, an $n$-qubit pure state \ket{\psi} that is SLOCC-equivalent to an $n$-qubit graph state \ket{g} obeys the relationship
\begin{equation}
\label{basic}
\ket{\psi} = S \ket{g} \text{,}
\end{equation}
where $S \in \rm{SLOCC} \brackets{2^{\it{n}}}$. Eq. (\ref{basic}) is a system of $2^n$ multivariate polynomial equations of degree $n$ in $4n$ unknowns (the four matrix elements of each of the $n$ local GL \brackets{2, \field{C}} operators. In fact, not only are each of the polynomial equations of degree $n$, the degree of each of the individual terms in each equation is also $n$. There are standard methods for determining whether there exists a solution to such a system of multivariate polynomial equations (for example, by computing a Gr\"{o}bner basis and examining the leading coefficients of each element)~\cite{Geddes1992}; however, these methods are generally very computationally expensive. In this section, the structure of graph states, and in particular their compact description using the stabilizer formalism, is exploited in order to give a different form of necessary and sufficient multivariate polynomial conditions for SLOCC-equivalence between \ket{\psi} and \ket{g}, generically having degree much lower than $n$. These conditions can then be examined, roughly speaking, in ascending order of degree, to determine SLOCC-equivalence. The benefits of this scheme are twofold. First, SLOCC-inequivalence can often be detected by examining the low-degree conditions, eliminating the need to look at high-degree multivariate polynomials. Second, the number of multivariate polynomials of degree $n$ that must be inspected is generally much lower than $2^n$.

The first relevant observation about graph states is a simple one, namely that the expectation value of any tensor product of single-qubit Pauli-$Z$ and identity operators vanishes for a graph state, except for the case of the identity operator itself.

\begin{lem}
\label{ortho}
Let \ket{g} be an $n$-qubit graph state corresponding to the underlying graph $G = \brackets{\mathcal{V},\mathcal{E}}$, having Pauli stabilizer $\Sigma(G)$. Then,
\begin{equation} 
\label{expectation}
\bra{g} \sigma Z_{\mbs{j(1)}} \ket{g} = \delta_{j,0}
\end{equation}
for all $\mbs{j} \in \bs{n}$ and for all $\sigma \in \Sigma(G)$.
\end{lem}

\begin{proof}
It is well-known that $\bra{g} Z_{\mbs{j(1)}} \ket{g} = \delta_{j,0}$ (see, for example, Ref.~\cite{Hein2006}). Since $\sigma$ is Hermitian, $\bra{g} = \bra{g} \sigma$ and thus,
\begin{equation}
\nonumber \bra{g} \sigma Z_{\mbs{j(1)}} \ket{g} = \delta_{j,0} \text{.}
\end{equation}
\end{proof}

Crucially, it turns out that the set of $4^n$ conditions of Eq.~(\ref{expectation}) uniquely specify the state \ket{g}, up to a global phase.

\begin{lem}
Let \ket{g'} be an $n$-qubit pure state and \ket{g} be an $n$-qubit graph state with Pauli stabilizer $\Sigma(G) = \set{\sigma_i}$. Then, 
\begin{equation}
\bra{g'} \sigma_i Z_{\mbs{j(1)}} \ket{g'} = \delta_{j,0} 
\end{equation}
for all $\sigma_i \in \Sigma(G)$ if and only if $\ket{g'} = \ket{g}$, up to a global phase.
\end{lem}

\begin{proof}
First, we prove the forward direction. Notice that since $\Sigma(G)$ is a stabilizer, all of its elements are commuting and share the same eigenbasis. In fact, it is clear that their mutual orthonormal eigenbasis is $\set{\ket{v_j} = Z_{\mbs{j(1)}} \ket{g}}$. We already know from Lemma~\ref{ortho} that these states are orthonormal. The fact that they are eigenstates of all of the stabilizer elements of \ket{g} can be seen by noting that all elements of the Pauli group either commute or anticommute, so $\sigma_i Z_{\mbs{j(1)}} \ket{g} = \pm Z_{\mbs{j(1)}} \sigma_i \ket{g} = \pm Z_{\mbs{j(1)}}  \ket{g}$. Note that all of the eigenvalues of the stabilizer elements corresponding to these eigenvectors are equal to $\pm 1$. Writing $\ket{g'} = \sum_j a_j \ket{v_j}$  and $\sigma_i = \sum_j \lambda^{(i)}_j \ketbra{v_j}{v_j}$, where the $\lambda^{(i)}_j = \pm 1$ and evaluating the expectation value of $\sigma_i$ in the state \ket{g'}, we find
\begin{eqnarray*}
\bra{g'} \sigma_i \ket{g'} & = & \displaystyle \sum_{j,k,l} \bra{v_j} a_j^* \lambda^{(i)}_k \ketbra{v_k}{v_k} a_l \ket{v_l} \\
                                           & = & \displaystyle \sum_{j,k,l} \lambda^{(i)}_k a_j^* a_l \braket{v_j}{v_k}   \braket{v_k} {v_l} \\
                                           & = & \displaystyle \sum_{j} \lambda^{(i)}_j \left| a_j \right|^2 \\
                                           & = & 1 \text{,}
\end{eqnarray*}
by assumption. In particular, for the case that $\sigma_i$ is the identity operator, all of the $\lambda^{(i)}_j$ are equal to 1, so 
\begin{equation}
\label{norm} \sum_{j} \left| a_j \right|^2 = 1 \text{.} 
\end{equation}
However, for any other stabilizer element, some of the $\lambda^{(i)}_j$ are equal to -1. The only eigenvector \ket{v_j} for which the 
corresponding eigenvalue is 1 for all of the stabilizer elements is the graph state itself, $\ket{v_0} = \ket{g}$. Thus, the only way Eq.~(\ref{norm}) can be satisfied for all stabilizer elements is if $|a_j| = \delta_{i,j}$, which in turn implies that $\ket{g'} = \ket{g}$, up to a possible global phase. So we have proven that $\ket{g'} = \ket{g}$ if $\bra{g'} \sigma_i \ket{g'} = 1$, which in light of Lemma~\ref{ortho} tells us that $\ket{g'} = \ket{g}$ if $\bra{g'} \sigma_i Z_{\mbs{j(1)}} \ket{g'} = \delta_{j,0}$, thereby completing the proof of the forward direction.

The reverse direction follows trivially from Lemma~\ref{ortho}.
\end{proof}

Similar conditions must be satisfied by SLOCC-transformed graph states, which can be determined by appropriately inserting resolutions of the identity $I = S S^{-1}$ into Eq.~(\ref{expectation}).

\begin{rem}
\label{sloccremark}
Let $\ket{\psi} = S \ket{g}$ and $\bra{\phi} = \bra{g} S^{-1}$, where $S \in \rm{SLOCC} \brackets{2^{\it{n}}}$ and \ket{g} is an $n$-qubit graph state with underlying graph $G = \brackets{\mathcal{V},\mathcal{E}}$. Then,
\begin{equation}
\label{conds1} 
\bra{\phi} \twiddle{Z}_{\mbs{j(1)}} \ket{\psi} = \delta_{j,0}
\end{equation}
for all $j \in \bs{n}$, with $\twiddle{Z}_{\mbs{j(1)}} = S Z_{\mbs{j(1)}} S^{-1}$.
\end{rem}

It should be noted that in general $\bra{\phi}$ does not describe a normalized vector, regardless of whether $\ket{\psi}$ was normalized. In principle, Eqs.~(\ref{conds1}) are conditions that must be satisfied by any \ket{\psi} that is SLOCC-equivalent to a graph state, and can therefore be used to test whether a given \ket{\psi} is inequivalent to a graph state. In practice, however, these conditions are not immediately useful as written. This is because we are not able to obtain \bra{\phi} from \ket{\psi}, as $S^{-1}$ (or equivalently $S$) is not known. This deficiency can be resolved by means of an observation relating \bra{\phi} to \bra{\psi}.

\begin{lem}
\label{rewritephi} Let $\ket{g}$ be an $n$-qubit graph state with underlying graph $G=\brackets{\mathcal{V},\mathcal{E}}$. Let $\ket{\psi} = S \ket{g}$ and $\bra{\phi} = \bra{g}S^{-1}$ for $S \in \rm{SLOCC} \brackets{2^{\it{n}}}$. Let $\sigma \in \pauli{n}$ be an element of the stabilizer for \ket{g}. Let $\ket{g^*} = \brackets{\ket{g}}^*$ denote the complex conjugate of \ket{g}. Then,
\begin{equation}
\bra{\phi} = \frac{1}{\det\brackets{S}} \bra{\psi^*} Y_\mathcal{V}SY_\mathcal{V} \sigma S^{-1}
\end{equation}
in any basis where $\ket{g}=\ket{g^*}$.
\end{lem}

\begin{proof}
It is easy to verify that
\begin{equation}
\label{inversetrick} S_i^{-1} = \frac{1}{\det{S_i}} Y_iS_i^TY_i \text{,}
\end{equation}
where $i \in \mathcal{V}$. This immediately implies that
\begin{equation}
S^{-1} = \frac{1}{\det{S}} Y_{\mathcal{V}}S^TY_{\mathcal{V}} \text{.}
\end{equation}
Therefore,
\begin{equation}
\ket{g} = S^{-1} \ket{\psi} = \frac{1}{\det{S}} Y_\mathcal{V}S^TY_\mathcal{V} \ket{\psi} \text{.}
\end{equation}
Taking the transpose and assuming we are working in a basis where $\ket{g} = \ket{g^*}$ (such as the computational basis) gives
\begin{equation}
\bra{g} = \frac{1}{\det{S}} \bra{\psi^*}Y_\mathcal{V}^T \brackets{S^T}^T Y_\mathcal{V}^T = \frac{1}{\det{S}} \bra{\psi^*}Y_\mathcal{V}SY_\mathcal{V} \text{.}
\end{equation}
Note that $\bra{\psi^*} = \ket{\psi^*}^{\dagger} = \ket{\psi}^T$. Right multiplying with $\sigma S^{-1}$ then yields
\begin{eqnarray*}
\bra{\phi} & = & \bra{g}S^{-1} \\
           & = & \bra{g} \sigma S^{-1} \\
           & = & \frac{1}{\det{S}} \bra{\psi^*} Y_\mathcal{V}SY_\mathcal{V} \sigma S^{-1} \text{.}
\end{eqnarray*}
\end{proof}

Lemma~\ref{rewritephi} and Remark~\ref{sloccremark} are all of the ingredients that are necessary for proving the main theorem, Theorem~\ref{mainthm}.
\begin{proof} \textbf{(of Theorem 1)} \\ 
Using Lemma~\ref{rewritephi}, Eq.~(\ref{conds1}) can be rewritten in the computational basis as
\begin{eqnarray*}
\bra{\phi} Z_{\mbs{j\brackets{1}}} \ket{\psi} 
& = & \frac{1}{\det\brackets{S}} \bra{\psi^*} Y_\mathcal{V} S Y_\mathcal{V} \sigma S^{-1} \twiddle{Z}_{\mbs{j\brackets{1}}} \ket{\psi} \nonumber \\ 
& = & \delta_{j,0}.
\end{eqnarray*}
Rearranging the expression above gives us the requisite form of the necessary and sufficient conditions, namely
\begin{equation}
\label{newconddefntemp} 
\bra{\psi^*} Y_\mathcal{V} S Y_\mathcal{V}  \twiddle{Z}_{\mbs{j\brackets{1}}} \ket{\psi} = \det\brackets{S} \delta_{j,0} \text{.}
\end{equation}
\\ 
\end{proof}

\section{Practical test of SLOCC-equivalence}
\label{sec:test}
Eqs.~(\ref{newconddefntemp}) constitute a set of necessary and sufficient conditions for SLOCC-equivalence between a graph state \ket{g} and a pure state \ket{\psi}. Thus, a given state \ket{\psi} can be tested for SLOCC-equivalence to \ket{g} simply by checking all of the conditions one at a time, setting the matrix elements of the location SLOCC-operators $S_i$ to be unknowns. As mentioned in Section~\ref{sec:introduction}, however, there are a few different types of conditions to be considered. First, the Category I conditions, those involving an odd number of unknown transformed Pauli operators for states with even numbers of qubits, and vice-versa, are automatically satisfied by all pure states and do not aid in the SLOCC-equivalence test, as described in Lemma~\ref{oddevenlem}. Here is the proof of that statement.
\begin{proof} \textbf{(of Lemma~\ref{oddevenlem})} \\
Using Eq.~(\ref{inversetrick}), we obtain
\begin{eqnarray*}
\bra{\psi^*} Y_{\mathcal{V}} \twiddle{Z}_J \ket{\psi} & = & \bra{\psi^*} Y_{\mathcal{V}} S Z_J S^{-1} \ket{\psi} \\
                                          & = & \bra{\psi^*} \brackets{S^T}^{-1} Y_{\mathcal{V}} Z_J S^{-1} \ket{\psi}~\text{det}\brackets{S} \text{.}
\end{eqnarray*}
Taking the transpose of the above expression and noting that $Y^T = -Y$ gives us
\begin{eqnarray*}
\bra{\psi^*} Y_{\mathcal{V}} \twiddle{Z}_J \ket{\psi} & = & \bra{\psi^*} \brackets{S^T}^{-1} Z_J Y_{\mathcal{V}} S^{-1} \ket{\psi}~\text{det}\brackets{S} \brackets{-1}^{|\mathcal{V}|} \text{.}
\end{eqnarray*}
Finally, using the anticommutativity of $Y$ and $Z$ yields
\begin{eqnarray*}
\bra{\psi^*} Y_{\mathcal{V}} \twiddle{Z}_J \ket{\psi} & = & \bra{\psi^*} \brackets{S^T}^{-1} Y_{\mathcal{V}} Z_J S^{-1} \ket{\psi}~\text{det}\brackets{S} \brackets{-1}^{|\mathcal{V} \setminus J|} \\
                                          & = & \bra{\psi^*} Y_{\mathcal{V}} \twiddle{Z}_J \ket{\psi}~\brackets{-1}^{|\mathcal{V} \setminus J|} \text{.}
\end{eqnarray*}
\end{proof}

Next, there are the Category II conditions, those that, while only being satisfied by SLOCC-transformed graph states, do not restrict the form of the SLOCC transformation, as given in Lemma~\ref{allzeroslemma}. This statement is proven below:

\begin{proof} \textbf{(of Lemma~\ref{allzeroslemma})}
The case for which $|\mathcal{J}|$ and $n$ have different parity is trivially covered by Lemma~\ref{oddevenlem}, so suppose that they have the same parity. Consider a set of 3 operators $\set{O_{\mathcal{J} \setminus i}X_i, O_{\mathcal{J} \setminus i}Y_i, O_{\mathcal{J} \setminus i} Z_i} \in \set{O^{(i)}_{\mathcal{J}}}$ that are identical except for the local operator acting on site $i \in \mathcal{J}$. Assume the conditions
\begin{eqnarray}
\label{allzerosX} \bra{\psi^*} Y_{\mathcal{V}} S_{\mathcal{V}}O_{\mathcal{J} \setminus i} X_i S^{-1}_{\mathcal{V}} \ket{\psi} & = & 0; \\
\label{allzerosY} \bra{\psi^*} Y_{\mathcal{V}} S_{\mathcal{V}}O_{\mathcal{J} \setminus i} Y_i S^{-1}_{\mathcal{V}} \ket{\psi} & = & 0; \\
\label{allzerosZ} \bra{\psi^*} Y_{\mathcal{V}} S_{\mathcal{V}}O_{\mathcal{J} \setminus i} Z_i S^{-1}_{\mathcal{V}} \ket{\psi} & = & 0
\end{eqnarray}
to be true. From Lemma~\ref{oddevenlem} we can infer that
\begin{equation}
\label{allzerosI} \bra{\psi^*} Y_{\mathcal{V}} SO_{\mathcal{J} \setminus i} I_iS^{-1} \ket{\psi} = 0. 
\end{equation}
Constructing an arbitrary linear combination of Eqs.~(\ref{allzerosX}),~(\ref{allzerosY}),~(\ref{allzerosZ}) and~(\ref{allzerosI}), we find that
\begin{equation}
\label{allzerossuper} \bra{\psi^*} Y_{\mathcal{V}} S_{\mathcal{V} \setminus i} O_{\mathcal{J} \setminus i} S^{-1}_{\mathcal{V} \setminus i} S_i \brackets{aX + bY +cZ + dI} S^{-1}_i  \ket{\psi} = 0 \text{,}
\end{equation}
where $a, b, c, d \in \field{C}$. But any arbitrary two-by-two matrix $M$ over the complex numbers can be written as
\begin{equation}
M = S_i \brackets{aX+bY+cZ+dI} S_i^{-1} \text{.}
\end{equation}
Therefore, Eq.~(\ref{allzerossuper}) is true for arbitrary $S$.
\end{proof}

The above results tell us how to go about formally testing for SLOCC-equivalence between \ket{\psi} and \ket{g}:
\begin{enumerate}
\item Construct formally all of the multivariate polynomial conditions of Eq.~(\ref{newconddefntemp}) for the cases in which $\mbs{j}=\mbs{00 \dots 0}$, i.e. for which the right hand sides equal 1, treating the matrix elements of the local SLOCC-operators $\set{S_i}$ as unknowns to be determined.
\item For each of the conditions obtained from the preceding step, construct all of the remaining conditions in which the unknown Pauli operator appearing has the same target (i.e. acts non-trivially on the same subset of the qubits), having right hand side zero. These are the Category III conditions.
\item The remaining conditions are all Category I or Category II and can be ignored, as they provide no information about the unknown SLOCC-operator $S$.
\item \label{laststep} Use the Category III conditions in increasing order of polynomial degree to solve for the unknown matrix elements of the $\set{S_i}$.
\end{enumerate}

Item~(\ref{laststep}) above may in practice be computationally difficult, so this SLOCC-equivalence test is not in general efficient. However, it is possible to detect SLOCC-inequivalence earlier than may otherwise have been possible, as there are conditions that do not depend on the elements of the $S_i$, and also conditions that do depend on these elements, but are of low degree (see the examples of Section~\ref{sec:introduction}). As a final comment, we note that for the case of graph states, the notions of SLOCC- and LU-equivalence coincide~\cite{VandenNest2004c}, so this scheme may provide a generically efficient means for detecting LU-inequivalence of two graph states.

\section{Constructing separable stabilizers for SLOCC-transformed graph states}
\label{sec:constructing}
In this section, we briefly show how a separable general stabilizer for a SLOCC-transformed $n$-qubit graph state $\ket{\psi} = S \ket{g}$ where $S \in \rm{SLOCC} \brackets{2^{\it{n}}}$ can be built constructively given the state vector. By general stabilizer, we mean a group of order $2^n$, each element of which fixes \ket{\psi}, but which is not necessarily a subgroup of the Pauli group \pauli{n}. By separable general stabilizer, we mean a general stabilizer whose elements can be expressed as tensor products of single-qubit (not necessarily Pauli) operations. First, we show how to construct the separable stabilizer (in the usual sense where the local operators are Pauli) for a graph state itself, given the state vector.

\begin{thm}
\label{sepstabtheorem} Let \ket{g} be an $n$-qubit graph state corresponding to an underlying graph $G = (\mathcal{V},\mathcal{E})$. Let $\ket{v_j} = Z_{\mbs{j\brackets{1}}} \ket{g}$ and $f_j = \ketbra{v_j}{v_j}$. Then, the set $\set{\sigma_i = \sum_j (-1)^{\mbs{i} \cdot \mbs{j}} f_j}$ is a separable Pauli stabilizer for \ket{g}.  
\end{thm}

\begin{proof}
Let $\ket{g} = \mathfrak{G} \ket{+}^{\otimes n}$, where $\mathfrak{G} = \prod_{(i,j) \in \mathcal{E}} CZ_{ij}$ is the product of controlled-$Z$ operators between all pairs of qubits having an edge between them. Then,
\begin{eqnarray*}
\sigma_i & = & \sum_j (-1)^{\mbs{i} \cdot \mbs{j}} f_j \\
         & = & \mathfrak{G} \sum_j (-1)^{\mbs{i} \cdot \mbs{j}} Z_{\mbs{j(1)}} \left( \ketbra{+^{\otimes n}}{+^{\otimes n}} \right)  Z_{\mbs{j(1)}} \mathfrak{G} \text{.}
\end{eqnarray*}
Assuming this works in the case where $\mathfrak{G}$ is the identity operator (i.e. the graph state has no edges), it will clearly work in the cases where the graph states do have edges, since the case for no edges gives a separable Pauli stabilizer, and $\mathfrak{G}$ is in the Clifford group. All that remains is to show that the construction works for the edgeless case, which can be done inductively. For the case of one qubit, we can easily verify that
\begin{eqnarray*}
\ketbra{+}{+} + \ketbra{-}{-} & = & I  \text{,} \\
\ketbra{+}{+} - \ketbra{-}{-} & = & X \text{.}
\end{eqnarray*}
These are manifestly the elements of the Pauli stabilizer for the one-qubit graph state. We can write this as the matrix equation
\begin{eqnarray}
\vec{s}_1 & = & \mtxtwotwo{1}{1}{1}{-1} \coltwo{\ketbra{+}{+}}{\ketbra{-}{-}} \\
          & = & \sqrt{2} H \coltwo{\ketbra{+}{+}}{\ketbra{-}{-}} \text{,}
\end{eqnarray}
where the components of the vector on the extreme right are two-by-two (density) matrices, $H$ is the Hadamard matrix
\begin{equation}
\nonumber H = \frac{1}{\sqrt{2}} \mtxtwotwo{1}{1}{1}{-1} \text{,}
\end{equation}
and $\vec{s}_1$ is also a vector of two-by-two matrices whose entries are the elements of the Pauli stabilizer for the one-qubit graph state. Now, for an inductive hypothesis, suppose that the $k$-qubit graph state has a corresponding stabilizer vector $\vec{s}_k$ given by
\begin{equation}
\vec{s}_k = 2^{\frac{k}{2}}H^{\otimes k} \coltwo{\ketbra{+}{+}}{\ketbra{-}{-}}^{\otimes k} \text{.}
\end{equation}
The vector of stabilizer elements for the $k+1$-qubit graph state corresponding to a graph with no edges is then clearly
\begin{eqnarray*}
\vec{s}_{k+1} & = & \left( \sqrt{2} H \coltwo{\ketbra{+}{+}}{\ketbra{-}{-}} \right) \otimes 2^{\frac{k}{2}}H^{\otimes k} \coltwo{\ketbra{+}{+}}{\ketbra{-}{-}}^{\otimes k} \\
              & = & 2^{\frac{k+1}{2}} H^{\otimes k+1} \coltwo{\ketbra{+}{+}}{\ketbra{-}{-}}^{\otimes k+1} \text{,}
\end{eqnarray*}
thereby proving the inductive hypothesis, and completing the proof.
\end{proof}

The above theorem shows how to find the stabilizer of a given graph state constructively, in terms of the $f_k$ matrices, which are obtained from the density matrix \ketbra{g}{g} conjugated by all possible $n$-fold tensor products of Pauli $Z$ operators and single-qubit identity operators. It is immediately apparent that if an operator $\sigma$ fixes \ket{g}, then the operator $S \sigma S^{-1}$ fixes $S \ket{g}$. Therefore, the following corollary, which tells how to constructively obtain the separable generalized stabilizer of a SLOCC transformed graph state $S\ket{g}$, holds immediately:

\begin{cor}
\label{sepgenstabtheorem} Let \ket{g} be an $n$-qubit graph state corresponding to an underlying graph $G = (\mathcal{V},\mathcal{E})$. Let $\ket{v_j} = Z_{\mbs{j\brackets{1}}} \ket{g}$ and $f_j = \ketbra{v_j}{v_j}$. Let $\ket{\psi} = S \ket{g}$ where $S \in \rm{SLOCC} \brackets{2^{\it{n}}}$. Then, the set $\set{\sigma_i = \sum_j (-1)^{\mbs{i} \cdot \mbs{j}} S f_j S^{-1}}$ is a separable general stabilizer for \ket{\psi}.  
\end{cor}

\section{Conclusions}
\label{sec:conclusion}
In this paper, we have provided a method for testing SLOCC-equivalence between a graph state and an arbitrary quantum pure state. The method offers two clear advantages over direct solution of the defining equation $\ket{\psi} = S \ket{g}$ for $S$: the multivariate polynomial equations to be solved are of generically much lower degree than those in the defining equation, and the number of equations of maximal degree to be solved is reduced. In particular, since the conditions to be satisfied can be arranged hierarchically in ascending order of polynomial degree, SLOCC-inequivalence between \ket{\psi} and \ket{g} can usually be detected without consideration of high-degree polynomial conditions. We have also provided a constructive method for determining the Pauli stabilizer of a given graph state; a method for constructing the generalized separable stabilizer for a SLOCC-transformed graph state follows naturally from here. 

The principal significance of this work is to aid in the determination of realistic systems, namely the ground states of physically realizable Hamiltonians, that are resource states for measurement-based quantum computing. Although it is well-known that such resources cannot arise exactly as the non-degenerate ground states of Hamiltonians involving only two-body interactions, it may nevertheless be possible for a state that is SLOCC-equivalent to a resource state to arise in such a context. The types of tasks that can be performed using such resource states (such as SLOCC-transformed cluster states) would in principle be the same as for the exact universal resource states themselves. In general, the scheme would involve POVMs rather than projective measurements, and the probability of success would in general be diminished from unity. The next step in this direction is to use the SLOCC-equivalence conditions generated in this work to examine some classes of physically realizable spin Hamiltonians to see if useful resources for MBQC arise as their ground states. 

A relevant question is whether any significant improvements can be made to the SLOCC-equivalence test as phrased in order to make it efficient in general, even in the worst case. Examining Eq.~\ref{newconddefntemp}, we see that in the case where $\mbs{j} = \mbs{00\dots0}$ and $\sigma_i$ acts as $Y$ in all but one position, the multivariate polynomial equation is reduced to a degree-2 equation. If for a particular graph there exists a transformation by means of local complementations on vertices to another graph~\cite{VandenNest2004} with a stabilizer element having $Y$ acting on all but qubit $k$, and this is possible for every choice of $k$, then the SLOCC-equivalence test will never require examination of any polynomial equations of degree higher than 2. It would be useful to identify classes of graphs for which such a local complementation scheme exists. 

It would also be of great interest to develop a scheme by which to test SLOCC-equivalence of pure states to the so-called Matrix Product States (MPS), many of which serve as universal resources for MBQC~\cite{Gross2007a}, and of which the graph states are a proper subset. Another question of interest would be to determine if the scheme can be simplified in the case where the given state \ket{\psi} is itself a graph state with known stabilizer, as this would allow for an efficient test of LU-equivalence between two graph states. These questions will be addressed in future work.

\section{Acknowledgements} The authors are grateful to Peter H\o yer for enlightening conversations and for simplifying the proofs of Lemmas~\ref{oddevenlem} and~\ref{rewritephi}. Funding for this project was provided by the Alberta Ingenuity Fund (AIF), the Natural Sciences and Engineering Research Council (NSERC), the Informatics Circle of Research Excellence (iCORE) and the Canada Foundation for Innovation (CFI).

\bibliographystyle{apsrev4-1}
\bibliography{succ6-arXiv}

\end{document}